\documentstyle[preprint,aps]{revtex}
\begin{document}
\draft
\title   {
          Restoration of isotropy on fractals.
         }
\author  {
Martin T.\ Barlow\cite{MTB}
} \address{
Department of Mathematics, University of British Columbia, Vancouver,
British Columbia V6T 1Z2, Canada
}
\author  {
Kumiko Hattori\cite{KH}
} \address{
Department of Mathematical Sciences, University of Tokyo, Komaba,
Tokyo 153, Japan
}
\author  {
Tetsuya Hattori\cite{TH}
} \address{
Faculty of Engineering, Utsunomiya University,
Ishii, Utsunomiya 321, Japan
}
\author  {
Hiroshi Watanabe\cite{HW}
} \address{
Department of Mathematics, Nippon Medical School, Kosugi,
Nakahara, Kawasaki 211, Japan
}
\maketitle
\begin{abstract}
We report a new type of restoration of {\em macroscopic isotropy
(homogenization)} in fractals with {\em microscopic anisotropy}.
The phenomenon is observed in various physical setups, including
diffusions, random walks, resistor networks, and Gaussian field theories.
The mechanism is unique in that it is absent in uniform media,
while universal in that it is observed in a wide class
of fractals.
\end{abstract}
%\narrowtext

In this letter, we report a new type of restoration of
macroscopic isotropy (homogenization) in fractals with microscopic anisotropy.
The phenomenon is unique in that it is absent in uniform media,
while universal in that it is observed in a wide class of fractals.
We suspect that the phenomenon is universal enough to be observed
experimentally, for example, in spin systems close to critical points
and various transport phenomena in fractal media.
We first discuss the Sierpi\'{n}ski gasket as an example of finitely
ramified fractals,
where the calculations can be performed explicitly.
We then turn to the Sierpi\'{n}ski carpet, an infinitely ramified fractal,
and report on rigorous results.
We conclude by discussing an intuitive picture of the mechanism.
Some results of numerical calculations are also presented.

We note that when we discuss ^^ isotropy' for a deterministic regular fractal,
we mean invariance with respect to (discrete) rotations
which respect the structure of the fractal.

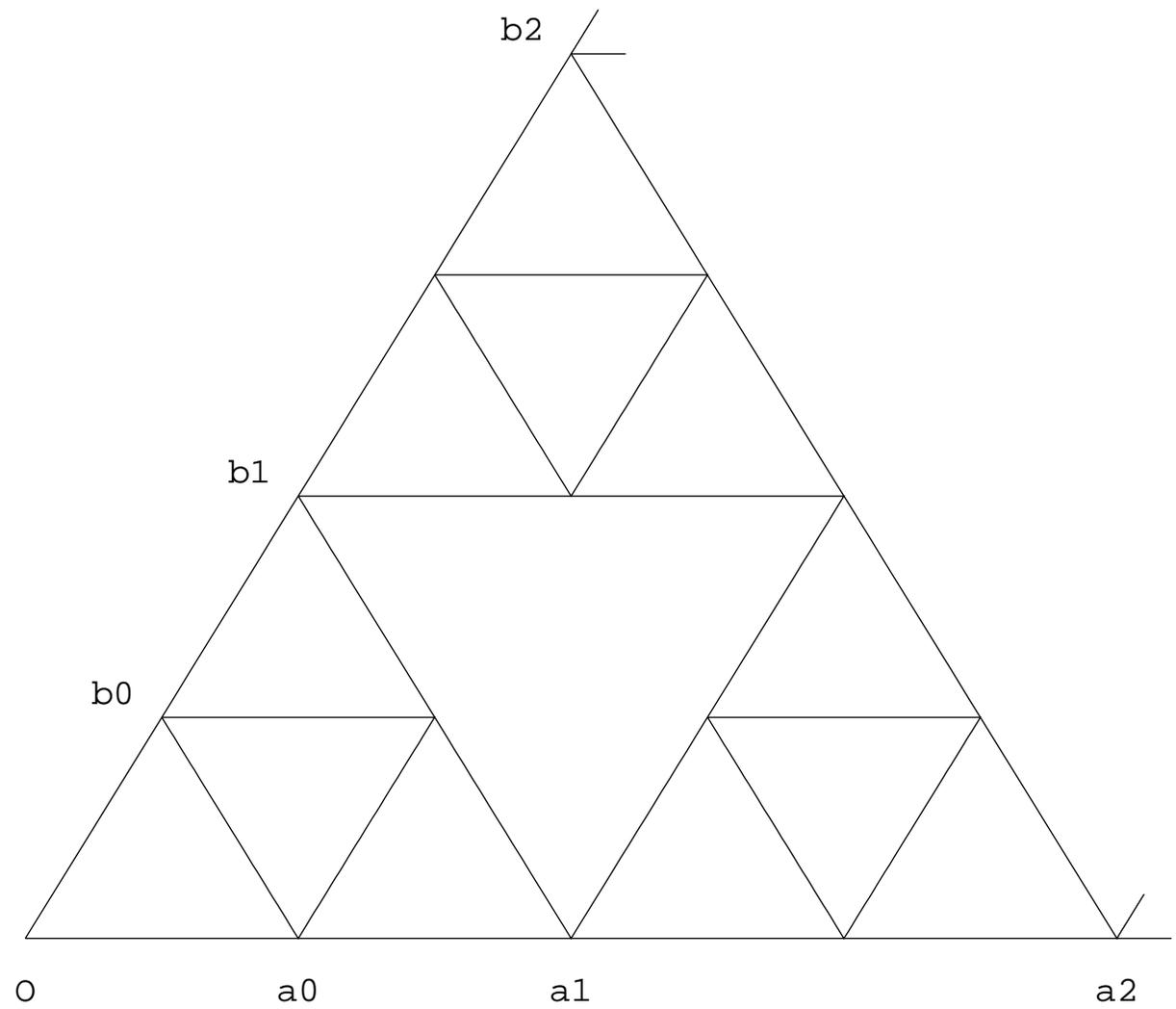
\begin{figure}
\caption{Pre-Sierpi\'{n}ski gasket.}
\label{fg:sg} %fig1.ps (psg.in)
\end{figure}

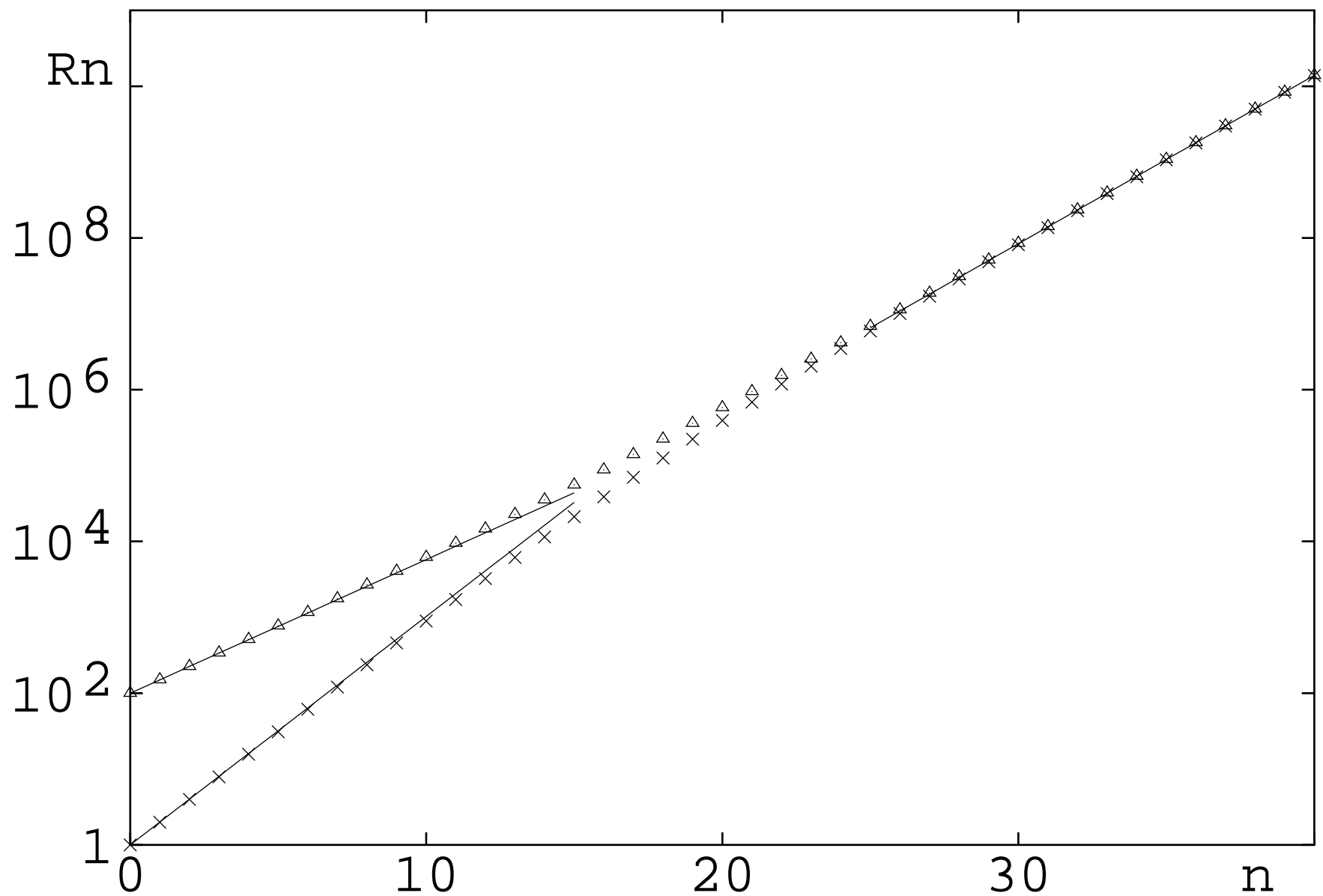
\begin{figure}
\caption{$R^x_n(r)$ (lower plots)  and $R^y_n(r)$ (upper plots)
on the pre-Sierpi\'{n}ski gasket for $r=100$.
The lines are the scaling predictions (\protect\ref{e:sganiso}) and
(\protect\ref{e:sgiso}). }
\label{fg:Rn} %fig2.ps=bf2.ps (sgh.in, sgh*.dat)
\end{figure}

\paragraph*{Resistor network on Sierpi\'{n}ski gasket.}
In order to illustrate the phenomenon of isotropy restoration, we
first concentrate on the simplest example of anisotropic resistor network
on the Sierpi\'{n}ski gasket, a typical finitely ramified fractal.
Let $n$ be a non-negative integer, and put $O=(0,0)$, $a_n=(2^n,0)$,
and $b_n=(2^{n-1},2^{n-1} \sqrt{3})$.
Consider the $n$-th generation of the (pre-)Sierpi\'{n}ski gasket,
which is a triangle $\triangle Oa_n b_n$
with self-similar internal structure composed of triangles of side length $1$,
as illustrated in (Fig.~\ref{fg:sg}).
Each internal vertex has $4$ bonds of unit length attached.
We associate a resistor of resistance $1$ with each bond parallel to
the $x$-axis, and a resistor of resistance $r>1$ with the remaining bonds.
By repeated use of the star--triangle relations ($Y$--$\Delta$ transforms),
this $n$-th level network can be reduced to a simple triangular network
(an effective network), with resistances $R^x_n(r)$ in the horizontal bond
$Oa_n$ and $R^y_n(r)$ in the bonds $Ob_n$ and $a_n b_n$.
By definition, $R^x_0(r)=1$ and $R^y_0(r)=r$.
Put
\begin{equation} \label{e:h} H_n(r) = R^y_n(r)/R^x_n(r) \,.  \end{equation}
$H_n(r)$ measures the effective anisotropy of $\triangle Oa_nb_n$ composed of
resistance elements with the basic (microscopic) anisotropy
parametrized by $r=H_0(r)$.
Using the star--triangle relations
we obtain the following recursion relations for $R^x_n$ and $R^y_n$:
\[ \begin{array}{c}
\displaystyle
R^x_{n+1}= \frac{2 R^x_n  R^y_n (2 R^x_n +3 R^y_n) (3 R^x_n + 2 R^y_n)}{
({R^x_n}^2 + 6 R^x_n R^y_n + 3 {R^y_n}^2) (R^x_n + 2 R^y_n)} \,, \\
\displaystyle
R^y_{n+1}=  \frac{R^y_n (2 R^x_n + 3 R^y_n)}{R^x_n + 2 R^y_n} \,.
\end{array} \]
We see from these formula that in the anisotropic regime
($H_n(r) \gg 1$), the effective resistances satisfy the scaling behavior
\begin{equation} \label{e:sganiso} R^x_{n+1}(r) \approx 2 R^x_n(r)\,,\ \
R^y_{n+1}(r) \approx (3/2) R^y_n(r)\,, \end{equation}
while in the isotropic regime ($H_n(r) \approx 1$),
we have
\begin{equation} \label{e:sgiso}
R^x_{n+1}(r) \approx R^y_{n+1}(r) \approx (5/3) R^x_n(r) \,. \end{equation}
We also see that $H_n(r)$ in (\ref{e:h}) satisfies
$H_{n+1}(r)^{-1} = f(H_n(r)^{-1})$, where
\begin{equation} \label{e:recsg} f(x)=(4 x+6 x^2)/(3+6 x +x^2) \,.
\end{equation}
In particular, we see the restoration of isotropy,
\begin{equation} \label{e:strong}
\lim_{n\to\infty} H_n(r)=1 \,. \end{equation}
Fig.~\ref{fg:Rn} gives the calculated behaviors of the effective resistances.
We see a clear signal of the two scaling regimes
(\ref{e:sganiso}) and (\ref{e:sgiso}).
Using (\ref{e:recsg}), we can calculate the rates of restoration of isotropy.
In the anisotropic regime, we have $H_{n+1}(r) \approx (3/4) H_n(r)$,
while in the isotropic regime, we have
$H_{n+1}(r)-1 \approx (4/5) (H_n(r)-1)$.
We can also calculate the scaling limit
$\displaystyle F(z) = \lim_{n\to\infty} f^n( (3/4)^n z )
= z - (3/2) z^2 + (39/14) z^3 + \cdots$,
where $f^n$ is the $n$-th iteration of $f$.
For large $r$ and large $n$
($1 \ll n < O(\log(r) / \log(4/3))$) we have
$ H_n^{-1}(r) \approx F( (4/3)^n /r ) $.
We can prove by standard methods using (\ref{e:recsg}) that the scaling limit
exists and that $F$ is complex analytic in a neighborhood of $z=0$.
\par
We can generalize the above consideration so that
the resistors parallel to $Ob_0$ and $a_0 b_0$ have different values.
If we denote the effective resistances parallel to $O a_0$, $O b_0$, $a_0 b_0$,
by $R^a_n$, $R^b_n$, $R^c_n$, respectively, we find
$ R^a_{n+1}=\frac{(4 K + R^a_n+R^b_n+R^c_n) R^a_n (R^b_n+R^c_n)}{
(K+R^b_n+R^c_n)(R^a_n+R^b_n+R^c_n)} $,
where $K=\frac{(R^a_n+R^b_n)(R^b_n+R^c_n)(R^c_n+R^a_n)}{
2(R^a_nR^b_n+R^b_nR^c_n+R^c_nR^a_n)}$.
Corresponding formula for $R^b_{n+1}$ and $R^c_{n+1}$ are obtained
by cyclic permutations of the suffixes.
Restoration of isotropy
$ \displaystyle \lim_{n\to\infty} R^b_n/R^a_n
= \lim_{n\to\infty} R^c_n/R^a_n = 1 $
can be proved in the generalized situation.
\par
Restoration of isotropy is not observed in uniform media.
To see this, consider a resistor network of regular square lattice,
whose horizontal (resp.\ vertical) bonds are resistors of resistance $1$
(resp.\ $r$).
The ratio of the effective resistances for $n \times n$ size network
in vertical direction to horizontal direction is easily seen
to be $r$, independently of $n$.
Thus the anisotropy for the resistor network of regular lattice is
independent of scale.
The restoration of isotropy which we observe on the Sierpi\'{n}ski gasket
is a feature absent on uniform media.

\paragraph*{Related models on Sierpi\'{n}ski gasket.}
We described restoration of isotropy in terms of resistor networks
\cite{W,BBprs}.
The phenomenon is also observed in various other physical setups,
including random walks and diffusions \cite{HHW93,H}
and Gaussian field theories \cite{HHW87}.
A related mathematical problem of the construction and uniqueness
of diffusions on the Sierpi\'{n}ski gasket is dealt with in \cite{O94}.
We also remark that there is another aspect in homogenization,
that a diffusion with microscopic irregularity restores macroscopic uniformity,
as studied in \cite{kumkus} for finitely ramified fractals.
This aspect, in contrast to what we deal with here, is not specific to
fractals and has been known in Euclidean spaces.
(For other related references in mathematics literature,
see the references in \cite{HNfl}.)

\paragraph*{Restoration of isotropy on Sierpi\'{n}ski carpet.}
The finite ramifiedness of the Sierpi\'{n}ski gasket implies that the
recursion relations are finite dimensional,
and the analysis can be made explicitly.
One might then wonder if the isotropy restoration we found above is
a special feature of models on finitely ramified fractals.
In \cite{BHHWreg} we have proved a mathematical theorem for a class of
infinitely ramified fractals,
which establishes that the isotropy restoration is a universal phenomenon.
\par
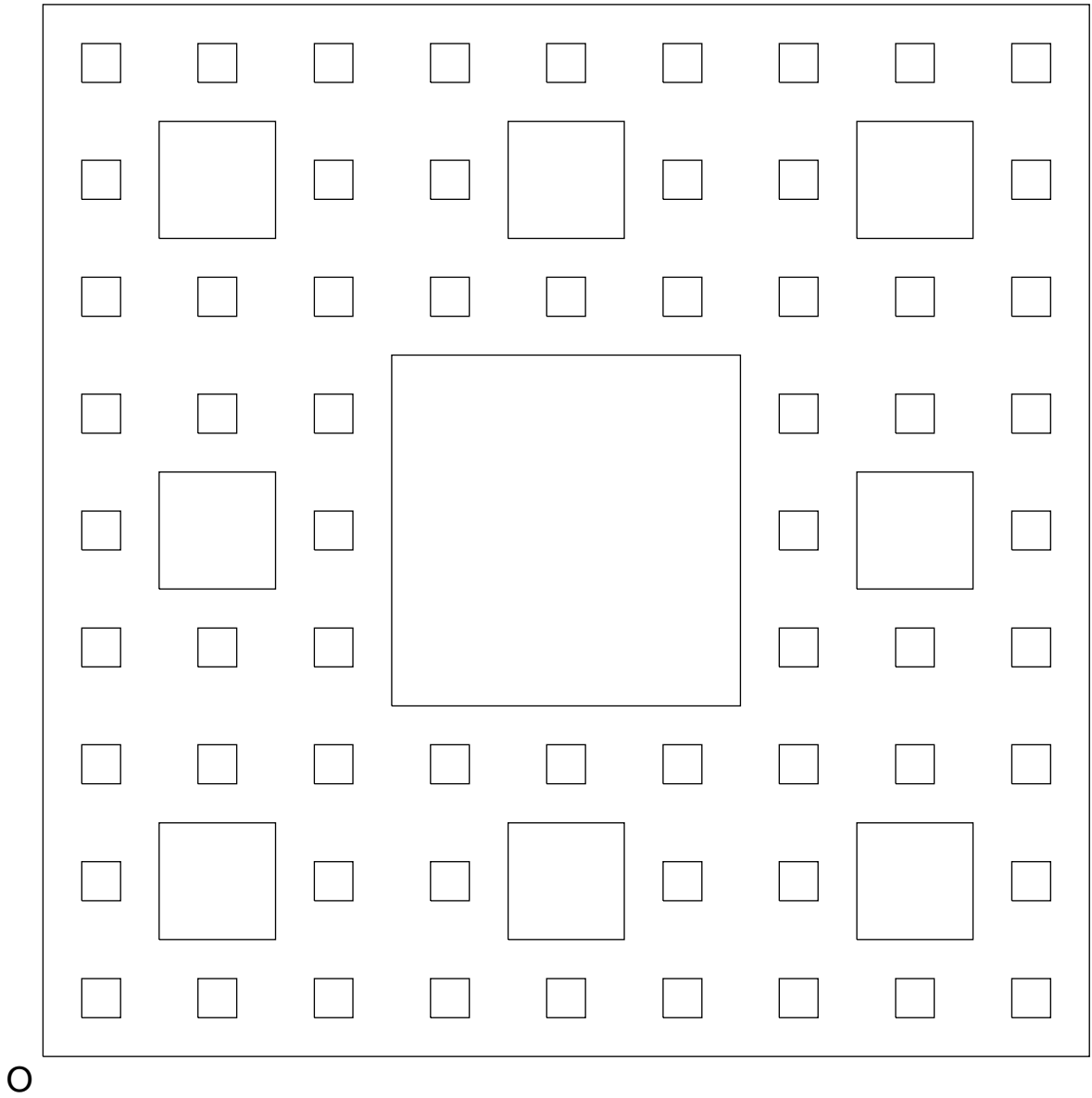
\begin{figure}
\caption{Pre-Sierpi\'{n}ski carpet $F_3$.}
\label{fg:sc} %fig3.ps=bf3.ps (sc.in, sc.dat)
\end{figure}

To state the result of \cite{BHHWreg},
let $n$ be a non-negative integer, and consider
the pre-Sierpi\'{n}ski carpet $F_n$,
which is a subset of a unit square $[0,1]\times[0,1]$
obtained by removing small squares recursively as for constructing
the Sierpi\'{n}ski carpet \cite{Sier},
until squares of side length $3^{-n}$ are reached, where we stop
so that smaller scale structures are absent (Fig.~\ref{fg:sc}).
Let $r>1$, and assume that $F_n$ is made of a material
with a uniform but anisotropic electrical resistivity, such that
for a unit square made of this material, the total resistance is $1$ in the
$x$-direction and $r$ in the $y$-direction, and the principal axes of the
resistivity tensor are parallel to the $x$ and $y$ axes.
Equivalently, we assume that
the energy dissipation rate per unit area for the
potential (voltage) distribution $v(x,y)$ is
$(\frac{\partial v}{\partial x})^2 + \frac{1}{r} \,
 (\frac{\partial v}{\partial y})^2$.
(Note that by linear transform in coordinate $y'=y \sqrt{r}$,
the formula becomes that of isotropic material.
Hence, in experimental situation, one may as well start with
a rectangle made of isotropic material, with rectangular holes.)
\par
We introduce the effective resistance $R^x_n(r)$ of $F_n$ in
$x$ direction, the resistance observed when
we apply voltage between two edges $x=0$ and $x=1$.
Likewise we define $R^y_n(r)$ and introduce the effective anisotropy $H_n(r)$,
as in (\ref{e:h}).
$H_0(r)=r$ parametrizes the anisotropy of the material composing $F_n$.
We can prove the following \cite{BHHWreg}.
\par\noindent {\em Theorem 1} ---
There is a finite constant $C\ge 1$, independent of $r$ and $n$,
such that for any initial anisotropy $r>0$,
we have the weak restoration of isotropy (weak homogenization)
in the sense that $1/C \le H_n(r) \le C$
holds for sufficiently large $n$.
(How large $n$ should be depends on the value of $r$.)
\smallskip\par
We believe that $C$ can be taken arbitrarily close to $1$,
as in (\ref{e:strong}), but this
is still beyond the reach of present mathematical techniques,
for the infinitely ramified fractals.
We emphasize that we have concrete rigorous results as Theorem 1,
in spite of the difficulties for the infinitely ramified fractals.
\par
Analogous results hold if we consider a cross-wire network $G_n$
defined by replacing each smallest size square of $F_n$ by a horizontal and
vertical cross-wire of four resistors (connected at the center of the square),
whose resistances are $1/2$ in
horizontal direction and $r/2$ in the vertical direction.
The results stated above for the board $F_n$ also hold for the network $G_n$.

\paragraph*{Ideas for a proof of the Theorem.}
Theorem 1 is proved by decomposing the problem into the isotropic regime
and the anisotropic regime.  For the isotropic regime, an extension
(to anisotropic case) of a deep renormalization group-type analysis of
effective resistance for the isotropic Sierpi\'{n}ski carpet \cite{BBprs,BBS}
is applied, while for the anisotropic case, renormalization group-type
picture in the neighborhood of degenerate fixed points \cite{HHW87,HHW93,H}
holds.
One of the key observations for the proof of Theorem 1 is that
if $H_n(r)$ is very large  (in the anisotropic regime),
then $H_n(r)$ follows a scaling behavior. We can prove
\par\noindent {\em Theorem 2} ---
The limits
$\displaystyle
\lim_{s\to\infty} s^{-1} \liminf_{n\to\infty} H_n((9/7)^n s)
= \lim_{s\to\infty} s^{-1} \limsup_{n\to \infty} H_n((9/7)^n s)$
exist.
\smallskip\par
This result says that while $s=(7/9)^n r$ and $n$ are large,
$H_n(r)$ decreases like $c\, (7/9)^n r$.
We can prove these Theorems by giving bounds controlling the
$n$ dependence of the effective resistances \cite{BHHWreg}.
Roughly speaking, we can show that in the anisotropic regime ($H_n(r) \gg 1$),
\begin{equation} \label{e:scaniso} R^x_{n+1}(r) \approx (3/2)\, R^x_n(r)\,,\ \
R^y_{n+1}(r) \approx (7/6)\, R^y_n(r)\,, \end{equation}
while in the isotropic regime ($H_n(r) \approx 1$),
$R^x_{n+1}(r) \approx \rho\, R^x_n(r)$, and
$R^y_{n+1}(r) \approx \rho\, R^y_n(r)$.
Here $\rho=1.25148 \pm 1\times 10^{-5}$
is the growth exponent for the effective resistance in the
isotropic case $r=1$ \cite{BBprs,BBS}.
\par
Based on these results,
we conjecture that (\ref{e:strong})
holds also for the Sierpi\'{n}ski carpet,
and that Fig.~\ref{fg:Rn} schematically
gives the behaviors of $R^x_n(r)$ and $R^y_n(r)$.

\paragraph*{Discussions.}
Our mathematical results are not very sharp numerically;
we can only say that $10^{-10} < H_n(r)< 10^{10}$, for large $n$.
Numerical calculations for the Sierpi\'{n}ski carpet may therefore be
of interest.
We give results for the resistor network $G_n$.
Obviously, $R^x_0(r)=1$ and $R^y_0(r)=r$.
It is not difficult to find
$R^x_1(r)= (3r+4)/(2r+3)$.
% R^x_1     R^y_1        H_1
% 1.400000  1.400000     1
% 1.478261  1.343750     10
% 1.497537  133.4437     100
% 1.499750  1333444.370  1000
The exact result for $n=2$ is
\[
R^x_2(r)=\frac{
324 r^8 + 3960 r^7 + 17169 r^6 + 37077 r^5 + 44639 r^4
+ 30842 r^3 + 11900 r^2 + 2325 r + 174}{144 r^8 + 1924 r^7 + 8850 r^6
+ 20052 r^5 + 25146 r^4 + 17976 r^3 + 7128 r^2 + 1422 r + 108}
%r=   1    r2x:=1.7934743202    r2y:=1.7934743202$
%r=  10    r2x:=2.085432551     r2y:=16.400548528$
%r= 100    r2x:=2.2257988183    r2y:=161.42304214$
%r=1000    r2x:=2.2474525852    r2y:=1611.4256332$
    \,. \]
Note that $R^y_n(r)= r\, R^x_n(1/r)$, with which we can calculate
$R^y_n(r)$ and $H_n(r)$ from these formula.
We have numerical results for $3 \le n \le 7$,
obtained using Gaussian relaxation method (Table~\ref{tb:1}).
We see that as $r$ is increased, the $n$ dependence of
$R^x_n(r)$ rapidly approach $(3/2)^n$,
and that for large $n$, those of $R^y_n(r)$ approach $c\,(7/6)^n$ with $c=6/5$.
These observations are consistent with (\ref{e:scaniso}),
implying scaling behavior in the anisotropic regime.
(Deviation from scaling of $R^y_n$ for small $n$ in the data can be explained
if we notice that we are calculating the network $G_n$ instead
of the board $F_n$.)
In particular, we see that for any value of $r>1$,
$R^y_n /R^x_n$ monotonically decreases as $n$ is increased,
which indicates the tendency of restoration of isotropy.
\par
We expect that the scaling limit
\[ z \lim_{n\to\infty} H_n((9/7)^n /z) = c + d\, z +\cdots  \]
exists, where $c$ is the limit in Theorem 2.
The data and the fact that $R^x_n(r)$ is a rational function of $r$
makes it possible to find an estimate
\[ R^x_n(0)=
\lim_{r\to\infty} r^{-1} R^y_n(r)=c\,(7/6)^n - 3^{-n}/5\]
with $c=6/5$.
Thus the constant term $c$ in the scaling function is determined.
We need more data to determine $d$, but
the calculations become rapidly time consuming as $n$ or $r$ is increased.
\par
Let us discuss general intuitive picture of the restoration of
isotropy, in terms of random walks \cite{HHW93,H}.
The fractals may be regarded to have obstacles or holes in the space,
when compared to uniform spaces.
Intuitively, a random walker that favors horizontal motion performs
a one-dimensional random walk between a pair of obstacles, and eventually
is forced to move in off-horizontal direction before he could move further
horizontally. There are obstacles of various sizes, separated by
distances of the same order as their sizes, hence globally, the random walker
is scattered almost isotropically.
On uniform media such as regular lattices or Euclidean spaces,
these obstacles are absent, hence the anisotropic walk keeps anisotropy
asymptotically.
\par
The Sierpi\'{n}ski gasket and the Sierpi\'{n}ski carpet have exact
self-similarity, and one may doubt
the ^^ extrapolation' to figures without exact self-similarity.
However, we can prove that the restoration of isotropy occurs
for anisotropic diffusions on the scale-irregular $abb$-gaskets,
a family of fractals which are
scale-irregular, i.e.\ do not have exact self-similarity \cite{H}.
These considerations suggest that the restoration of isotropy is to be
observed on a wide class of random media.
For example, numerical calculations on the percolation clusters
may provide interesting observations.
\smallskip\par
T.~Hattori wishes to thank Hal Tasaki for his interest in the present work,
for valuable comments, and above all,
for suggesting to write a letter on the subject.
The research of T.~Hattori is supported in part by a Grant-in-Aid for General
Scientific Research from the Ministry of Education, Science and Culture.

\begin{table}
\caption{Effective resistances $R^x_n(r)$ and $R^y_n(r)$
for the pre-Sierpi\'{n}ski carpet network $G_n$. }
\label{tb:1}
\begin{tabular}{|c|c|c|c|c|c|c|c|c|c|c|} \hline &\multicolumn{2}{c|}{$r=10$}
  &\multicolumn{2}{c|}{$r=100$} &\multicolumn{2}{c|}{$r=1000$}
&\multicolumn{2}{c|}{$r=10000$} &\multicolumn{2}{c|}{$r=100000$} \\ \hline
$n$ & $R^x_n(r)$ & $R^y_n(r)$
    & $R^x_n(r)$ & $R^y_n(r)$ & $R^x_n(r)$ & $R^y_n(r)$
    & $R^x_n(r)$ & $R^y_n(r)$ & $R^x_n(r)$ & $R^y_n(r)$
\\ \hline
$3$ & $2.831057$ & $19.64149$
    & $3.238145$ & $190.6445$  & $3.356806$ & $1899.017$
    & $3.373110$ & $18982.35$  & $3.374810$ & $189815.7$
\\ \hline
$4$ & $3.798415$ & $23.47825$
    & $4.614455$ & $224.0274$  & $4.963201$ & $2223.085$
    & $5.049858$ & $22209.29$  & $5.061194$ & $222070.3$
\\ \hline
$5$ & $5.070868$ & $28.10055$
    & $6.524220$ & $263.1750$  & $7.258880$ & $2598.702$
    & $7.524180$ & $25934.86$  & $7.585124$ &
\\ \hline
$6$ & $6.742934$ & $33.69136$
    & $9.185975$ & $309.3891$  & $10.56635$ & $3037.488$
    & $11.14879$ & $30272.61$  & $11.34244$ &
\\ \hline
$7$ & $8.933314$ & $40.46672$
    & $12.88375$ & $364.0724$  & $15.34037$
    &\multicolumn{5}{c|}{}
\\ \hline
\end{tabular}
\end{table}

\end{document}